\documentclass[journal,10pt,twocolumn,letterpaper]{IEEEtran}

\usepackage[utf8]{inputenc}
\usepackage[T1]{fontenc}

\usepackage[english]{babel}
\usepackage{amsbsy,amsmath,amsfonts,amssymb,amsthm}
\usepackage{mathtools}
\usepackage{textcomp} %degrees centigrade symbols, euros, etc. 

\usepackage[noadjust]{cite}
\usepackage{bm,cases,url,verbatim} 
\usepackage{booktabs}
\usepackage{centernot} % prints the sym­bol \not on the fol­low­ing ar­gu­ment

\newcommand{\subparagraph}{}

\newcommand{\argmin}{\operatornamewithlimits{argmin}}

\begingroup
\catcode`\#=11

\endgroup

\usepackage{graphicx,xcolor,float,dblfloatfix}
\usepackage{psfrag}

\usepackage{caption}
\usepackage{subcaption}

\newtheorem{thm}{Theorem}

\newtheorem{lemma}{Lemma}

\newcommand{\subalign}[1]{%
  \vcenter{%
    \Let@ \restore@math@cr \default@tag
    \baselineskip\fontdimen10 \scriptfont\tw@
    \advance\baselineskip\fontdimen12 \scriptfont\tw@
    \lineskip\thr@@\fontdimen8 \scriptfont\thr@@
    \lineskiplimit\lineskip
    \ialign{\hfil$\m@th\textstyle##$&$\m@th\textstyle{}##$\crcr
      #1\crcr
    }%
  }
}

\usepackage{epstopdf}

\graphicspath{%
{./figures/}
}

\usepackage{tikz}
\usetikzlibrary{arrows, decorations.markings}

\usepackage[flushleft]{threeparttable}

\usepackage{color}
\newcommand{\CP}[1]{{\color{cyan} \emph{CP: }#1}}

\newcommand{\ttx}{{\rm tx}}
\newcommand{\pr}{{\rm P}} 
\newcommand{\prr}{{\rm Pr}} 
\newcommand{\out}{{\rm out}} 
\newcommand{\fl}{{\rm full}}

\begin{document}

\title{Minimizing Data Distortion of Periodically Reporting IoT Devices with Energy Harvesting}

\author{\IEEEauthorblockN{Chiara Pielli\IEEEauthorrefmark{1}, \v{C}edomir~Stefanovi\'c\IEEEauthorrefmark{2}, Petar~Popovski\IEEEauthorrefmark{2}and Michele~Zorzi\IEEEauthorrefmark{1}} \\
\IEEEauthorblockA{\IEEEauthorrefmark{1}Department of Information Engineering, University of Padova, Padova, Italy\\
\IEEEauthorrefmark{2}Department of Electronic Systems, Aalborg University, Aalborg, Denmark\\
Email: piellich@dei.unipd.it, cs@es.aau.dk, petarp@es.aau.dk, zorzi@dei.unipd.it}}

\maketitle

\begin{abstract}
Energy harvesting is a promising technology for the Internet of Things (IoT) towards the goal of self-sustainability of the involved devices. 
However, the intermittent and unreliable nature of the harvested energy demands an intelligent management of devices' operation in order to ensure a sustained performance of the IoT application. 
In this work, we address the problem of maximizing the quality of the reported data under the constraints of energy harvesting, energy consumption and communication channel impairments.
Specifically, we propose an energy-aware joint source-channel coding scheme that minimizes the expected data distortion, for realistic models of energy generation and of the energy spent by the device to process the data, when the communication is performed over a Rayleigh fading channel.
The performance of the scheme is optimized by means of a Markov Decision Process framework.
\end{abstract}

\section{Introduction} \label{sec:intro}

A lion's share of IoT applications involves scenarios in which a multitude of devices, i.e., sensors, periodically report a small amount of data about some monitored phenomena.
A major challenge in this regard is to ensure uninterrupted service with minimal device maintenance.
Specifically, a typical IoT device is expected to operate for very long periods without human intervention, achieving a minimum of 10 years of battery (i.e., device) operation in such IoT applications \cite{nokia, ericsson, huawei}.
Energy Harvesting (EH) is a promising technology that can foster the required self-sustainability of the devices and, thus, of the IoT applications.

On the other hand, the EH process is stochastic in nature, requiring rethinking and redesign of the operation with respect to standard approaches employed for battery-operated devices.
%A general overview of recent advances in wireless communications with EH is presented in \cite{Uetal2015}.
Specifically, in this work we consider a monitoring application where a sensor node powered by renewable energy sources periodically sends its measurements to a data gathering point. 
%Before transmission, it performs processing operations: lossy compression to remove the intrinsic data redundancy, and encoding to increase the robustness against channel performance. The goal is to guarantee self sufficiency of the network while maximizing the reconstruction fidelity at the receiver, which depends on the tradeoff between compression accuracy and probability of outage. 
Before transmission, it performs processing operations, e.g., lossy joint source-channel coding that maximizes the reconstruction fidelity at the receiver, under the constraints omposed by the EH process.
In particular, the reconstruction fidelity depends on the tradeoff between source compression accuracy and robustness against the channel impairments, and is represented in terms of data distortion.

The effect of packet losses on data distortion in standard communication scenarios, where there are no energy constraints, has already been treated, e.g., in~\cite{Zachariadis}, where the authors investigate erasure and scalable codes for a Gaussian source. Several works (e.g.,~\cite{etemadi}) analyze layered transmission schemes, where the source is coded in superimposed layers, and each layer successively refines the description of the previous one, but is transmitted with a higher coding rate (i.e., subject to larger outage probability).  
Often, the distortion exponent is adopted as the performance metric for the end-to-end distortion~\cite{Laneman,aguerri}. Nevertheless, the distortion exponent is meaningful only for the high SNR regime, which is not dominant in IoT scenarios.

A general overview of recent advances in wireless communications with EH is presented in \cite{Uetal2015}. Further, distortion minimization for EH sensor nodes has been studied in several works.
In~\cite{bhat}, the tradeoff between quantization and transmission energy in the presence of EH is analyzed, but the effect of packet losses on the received data quality is not taken into account.
The problem of energy allocation between processing and transmission is also studied in~\cite{Castiglione}, in a model similar to our own, but where, rather than minimizing the long-term distortion, the authors aim at guaranteeing a minimum average distortion while maintaining the data queue stable. 
In~\cite{motlagh}, the authors study the achievable distortion when the energy buffer may have some leakage and the transmitting devices jointly perform source-channel coding in the presence of Gaussian and binary sources.

The work closest to ours can be found in~\cite{zordan}, where a sensor node employs an \emph{on-line} transmission policy that maximizes the long-term average quality of the transmitted packets. Specifically, the transmitter can decide the degree of lossy compression and the transmission power, and the optimal transmission strategy is obtained through the use of Markov Decision Processes (MDPs).
The framework employed in this paper bears similarities to the one from~\cite{zordan}, with the following important differences: (i) in addition to source coding, we also consider channel coding, leveraging on the results of finite-length information theory; (ii) we do not perform power control, which requires Channel State Information (CSI) at the transmitter and thus implies an additional cost or overhead, and may be available only with a delay; and (iii) the actual distortion at the receiver is influenced not only by the source processing procedure, but also by the channel outage probability.

The problem considered in the paper is solved through decompositon into two nested optimization processes that respectively address the rate-distortion tradeoff and the energy management. A similar approach has been followed in~\cite{bui}, where the goal is to guarantee energy self-sufficiency to a multi-hop network by adapting the nodes' duty cycle and the information generation rate. Again, the framework is modeled by means of an MDP; in fact MDPs are widely used to address energy management policies, because they represent an appealing solution to optimize some long-term utilities in the presence of stochastic EH~\cite{tutorial}.
The nested optimization allows us to determine an online optimal policy: although solving an MDP may require some computational time, the optimal solution can be precomputed and stored, such that during its operation the node decides on its action with a simple table look-up.
%We show that the optimal policy is non-decreasing in the battery state.
We also note that the considered framework is based on realistic models of the EH process and of the distortion achievable by compressing environmental signals with practical algorithms, including a thorough energy consumption model.

The rest of the paper is organized as follows. The system model and the problem formulation are described in Sections~\ref{sec:model} and~\ref{sec:problem}, respectively. The rate-distortion tradeoff is analyzed in Section~\ref{sec:RDP}, whereas Section~\ref{sec:EMP} studies the energy management policy. Finally, we provide and discuss some exemplary numerical results in Section~\ref{sec:results} and draw the conclusions in Section~\ref{sec:conclu}.

\section{System model} \label{sec:model}

We consider an IoT device that generates packets periodically and communicates wirelessly with a data collector. 
Time is divided into slots of predefined duration $T$. % $n \in \mathbb{N}$ corresponds to the time interval $[t_n,t_{n+1})$ and has,
Each time the device generates a packet, it has access to a slot reserved for its transmission.
In terms of the classification from \cite{Uetal2015}, the considered scenario can be categorized as an on-line energy management with a reward maximization and with a perfect knowledge of the state of the energy buffer, for a single device case. 

\begin{figure}[t]
  \centering
  \includegraphics[width=0.95\columnwidth]{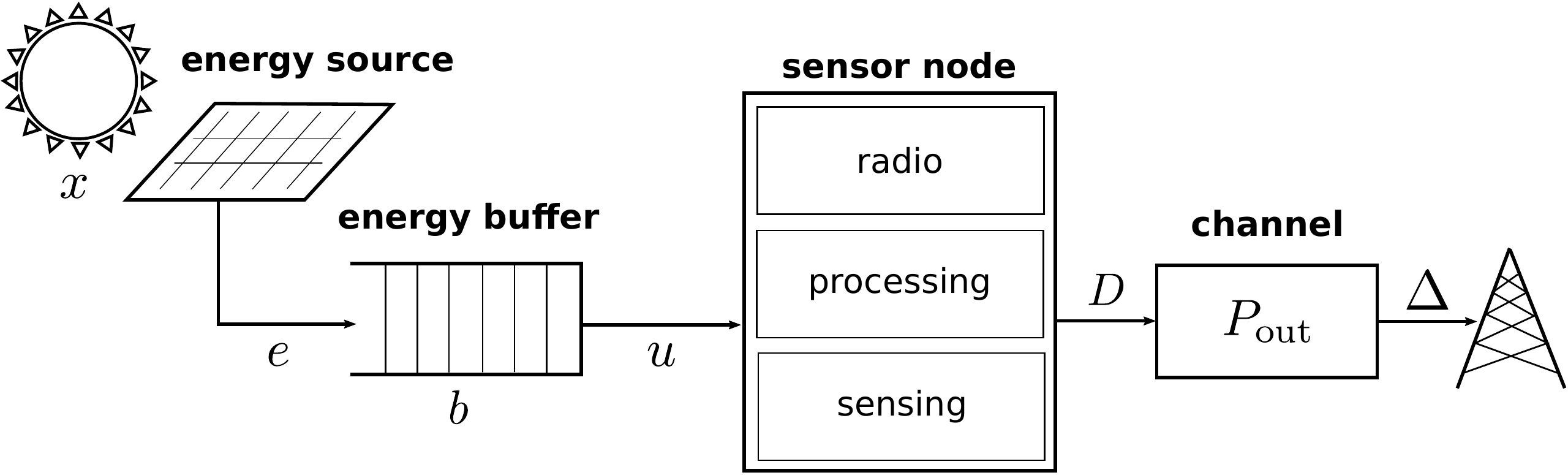}
  \caption{System model.}
  \label{fig:node_model}
\end{figure}

Fig.~\ref{fig:node_model} shows the dynamics of the sensor node:
 some energy-scavenging circuitry allows the device to harvest energy, which is stored in a buffer and used for sensing, processing and transmission operations. In the following, we thoroughly describe all the components of such model.
    
% \begin{figure}[h]
% \centering
% \def\svgwidth{.95\columnwidth}
% \input{figures/PROVA.pdf}
% \caption{System model.} 
% \label{fig:node_model}
% \end{figure}

\subsection{Data Processing and Transmission Model} \label{subsec:data}
The sensor node periodically senses the surrounding environment. In each time slot the device collects a constant number of readings for a total data size of $L_0$ bits per slot.
In the following, we consider a typical slot when defining quantities of interest; in general, we will introduce subscripts referring to slots only when needed. The sensor node may exploit source coding; in particular, we consider lossy compression algorithms, and the resulting packet size after compression is $L(k) = k/m \, L_0$, where $m\in\mathbb{N}$, $k \in \{0,1,\dots,m\}$. The device decides upon $k$ and thus on the compression ratio $k/m \in [0,1]$.
Lossy compression reduces the volume of bits of information to send, but at the cost of a poorer accuracy in the signal representation, as it introduces a distortion $D(k)$ that depends on the compression ratio.

The slot duration $T$ defines the maximum number of bits that can fit into a slot, namely $S=T/T_{\rm b}$, where $T_{\rm b}$ is the (fixed) bit duration. Hence, when deciding upon the degree of source compression $k$, the device also selects the coding rate $R(k)=L(k)/S$ at the same time.
We assume that no CSI is available at the transmitter%TODO: dire qualcosa estimate etc
, which, therefore, does not perform power control and, whenever transmitting, uses a constant transmission power $P_\ttx$.
Depending on $R(k)$ and on the actual channel conditions, the packet may not be correctly received with an outage probability $\pr_\out(k)$. 
Since we consider a monitoring scenario where the traffic pattern is reasonably known in advance, %TODO dirlo nell'intro
we assume that the slot duration is allotted in such a way that the device can send all the gathered information within the slot, thus $L_0\le S$.

Both the distortion function and the outage probability will be characterized in Section~\ref{sec:RDP}.

\subsection{Energy Harvesting Model} \label{subsec:eh}

We assume that the device is capable of collecting energy from the environment, and that this energy supply exhibits time correlation. 
Our approach follows the works based on realistic data~\cite{Ku,solarstat} that assess the goodness of Markov Chains (MCs) to model time-dependent sources.

The energy source dynamics are modeled by means of a 2-state MC,  and the number $e_n$ of energy quanta harvested in slot $n$ depends on the source state $x_n \in \mathcal{X} = \{0,1\}$. Based on the realistic model of~\cite{solarstat}, we assume that, when $x_n=0$, the source is in a ``bad'' state and $e_n=0$ with probability 1, otherwise, the source is in a ``good'' state and the energy inflow follows a truncated discrete normal distribution, i.e., $e_n \sim \mathcal{N}(\mu, \sigma^2)$ in the discrete interval $\{1,\dots\,E\}$. 

We consider a \emph{harvest-store-use} protocol: the energy scavenged in slot $n$ is first stored in a battery of finite size $B$ and then used from slot $n+1$ onwards.
In slot $n$, the battery state is $b_n \in \{0,\dots,B\}$, and evolves as follows:~
\begin{align}
 b_{n+1} & = \min\left\{b_n + e_n - u_n,\, B\right\} \triangleq \left(b_n + e_n - u_n\right)^\dagger, \label{eq:buffer}
\end{align}

\noindent where $u_n\ge 0$ is the energy used in slot $n$. The energy causality principle entails that:~
\begin{equation} \label{eq:en_causality}
 u_n \le b_n
\end{equation}
in all slots.
We assume that the sensor node can reliably estimate the status of its energy buffer.

\subsection{Energy Consumption Model}   \label{subsec:en_consumption}
%TODO qualche citazione di paper su energy consumption
We consider the following sources of energy consumption.

\emph{\textbf{Data processing}.}
To model the energy consumed by compression, we leverage on the results of~\cite{zordan_lossy}, where the authors map the number of arithmetic operations required to process the original signal into the corresponding energy demand:~
 \begin{align} \label{eq:e_processing}
   E_P(k)  = 
   \begin{cases}              
      E_0 \, L_0 \, N_P(k) & \text{if }  1\le k \le m-1\\
      0 & \text{if } k=0,m
   \end{cases}  
 \end{align}
 \noindent where $E_0$ models the energy spent by the micro-controller of the node in a clock cycle, and $N_P(k)$ is the number of clock cycles required by the compression algorithm per uncompressed bit of the input signal, which depends on the compression ratio. If the packet is not compressed ($k=m$) or discarded ($k=0$), no energy is consumed.
 
 Different compression algorithms entail different shapes of $N_P(k)$. In particular, there are some techniques for which function $N_P(k)$ is increasing and concave in $k$\footnote{Although this may appear as a counterintuitive result, it is due to practical implementation details. We refer the reader to~\cite{zordan_lossy} for a detailed explanation.}, e.g., the Lightweight Temporal Compression (LTC) and the Fourier-based Low Pass Filter (DCT-LPF) algorithms.  
 We consider the LTC algorithm, whose corresponding energy consumption turns out to be linear in the compression ratio: $N_P(k) = \alpha_P \frac{k}{m} + \beta_P,\, k\!\in\!\{1,\dots,m-1\}$,  with $\alpha_P, \beta_P > 0$~\cite{zordan_lossy}.
%   \begin{align} \label{eq:e_processing}
%    E_P(k)  = 
%    \begin{cases}              
%       E_0 \, L_0 (\alpha_P \frac{k}{m} + \beta_P) & \text{if }  1\le k \le m-1\\
%       0 & \text{if } k=0,m
%    \end{cases}
%  \end{align}
%  
%  \noindent with $\alpha_P, \beta_P > 0$.
 
\emph{\textbf{Data transmission}.}
The energy needed to transmit a signal with constant power $P_\ttx$ for a time window of length $T$ is:~
\begin{align}
 E_{TX}(k) = \frac{P_\ttx \,T}{\eta_A}\cdot\chi_{\{k>0\}},  \label{eq:e_tx}
\end{align}

\noindent where %the numerator represents the energy of the radio signal transmitted over the air, and 
$\eta_{A} \in (0,1]$ is a constant term that models the efficiency of the power amplifier of the antenna. The indicator function $\chi_{\{k>0\}}$, equal to 1 if $k\!>\!0$ and to zero otherwise, ensures that no energy is consumed if the packet is discarded.

\emph{\textbf{Other operations and circuitry costs}.}
We define the circuitry energy consumption in a slot as:~
\begin{align}
 E_C(k) =  \beta_S + \beta_C + \left(\beta_E + \mathcal{E}_C \,T\right)\cdot\chi_{\{k>0\}},  \label{eq:e_circuitry}
\end{align}

\noindent where $\beta_S$ and $\beta_E$, respectively, represent the constant sensing and encoding costs, $\beta_C$ accounts for switching between the node's operating modes and the maintenance of synchronization with the receiver, and the last term models the additional energy cost incurred by a transmission, where $\mathcal{E}_C$ is a circuitry energy rate. 
We highlight that, typically, the energy demanded by the channel encoding procedure is so small as to be negligible.

\medskip
The total energy consumption of the node is given by the sum of the contributions of Eqs.~\eqref{eq:e_processing}-\eqref{eq:e_circuitry}:~
\begin{equation} \label{eq:energy_used}
  E_{\rm used}(k) = E_P(k) + E_{TX}(k) + E_C(k),
\end{equation}

\noindent which depends solely on $k$. 
%We remark that the transmitter does not adapt the transmission power because of the lack of CSI, and thus both $E_{TX}$ and $E_C$ represent constant contributions in this scenario.

\section{Problem Formulation} \label{sec:problem}

Our goal is to minimize the long-term average distortion of the data transmitted by the sensor node while guaranteeing the self-sufficiency of the network.
The optimal distortion point is selected according to a joint source-channel coding policy: the node has to choose the number of bits per symbol to transmit, which implies to decide both the degree of lossy compression at the source and the error correction coding rate. The identification of such a working point is also driven by the energy dynamics: the node is powered through EH, hence the energy income is intermittent and erratic, and it is crucial to design an intelligent scheme to manage the available energy.

To deal with the two aspects of distortion and energy, we split the problem into two intertwined subproblems.

\begin{itemize}
 \item \emph{Rate-Distortion Problem (RDP)}: it focuses on the rate-distortion issue, and assumes to have a given amount of energy available to accomplish its goal. RDP will be discussed in Section~\ref{sec:RDP}.
 
 \item \emph{Energy Management Problem (EMP)}: this module dynamically decides the energy to use in each slot with the ultimate objective of ensuring long-term, uninterrupted operation. %The energy management strategy is computed given the energy consumption model (see Section~\ref{subsec:en_consumption}), the state of the energy buffer, and a statistical characterization of the energy inflow (see Section~\ref{subsec:eh}).
 This problem will be formulated and solved in Section~\ref{sec:EMP}.
\end{itemize}

The two subproblems are tightly coupled: RDP selects the optimal operating point on the rate-distortion curve as a function of the available energy, whereas EMP decides how this energy varies depending on the battery state and the statistical knowledge of the EH process. In this way, the node is assured to be self-sufficient in an energetic sense and, under this operating condition, the long-term average distortion is minimized. The combined problem is solved by means of an MDP, as described in Section~\ref{sec:EMP}.

Notice that the separability of RDP and EMP in the overall optimization process leads to optimality because, once we decide the energy to be spent from the energy buffer, we focus on the QoS and choose the action that minimizes the distortion and the overall optimization is guaranteed by solving the MDP.

\section{Rate-Distortion Analysis} \label{sec:RDP}

In this section, we determine the optimal point in the rate-distortion tradeoff when the available energy budget is given.
The actual distortion at the receiver is influenced by both the lossy compression performed at the source and the errors introduced by the channel. More specifically, we consider the packet to be completely lost when in outage, which corresponds to the maximum distortion level.

We first proceed to characterize the distortion due to source coding and the outage probability as a function of the rate, and then explain how to derive the solution of RDP.
%Since RDP focuses on a single slot, we will omit the dependence on the slot index $n$ throughout this section, unless misleading.

\subsection{Distortion due to compression} \label{subsec:dist}
As described in Section~\ref{subsec:data}, in each slot the device can compress its readings through a lossy compression algorithm, thereby achieving a lower rate $R(k)$, but, at the same time, introducing a certain degree of distortion.
In the literature, there exist closed-form expressions for the rate-distortion curves for idealized compression techniques operating on Gaussian information sources, but for practical algorithms such curves are generally obtained experimentally. 
% TODO\CP{literature}

With the aim of modeling a more realistic scenario, we leveraged on the work in~\cite{zordan_lossy} to derive a mathematical fit of rate-distortion curves that were obtained experimentally:~%. We ended up with the following expression:~
\begin{equation} \label{eq:dist}
D(k) = 
 \begin{cases}
  b \left( \left(\frac{k}{m} \right)^{-a} - 1\right) \quad &\text{if}\; k > 0\\
  D_{\fl} \quad & \text{if} \; k = 0,
  \end{cases}
\end{equation}

\noindent where $b\!>\!0$, $0\!<a\!<\!1$, and $D_\fl$ is the maximum distortion, reached when the data is not even transmitted (extreme case of compression ratio equal to 0). 
\begin{comment}
The distortion metric we adopt is the maximum absolute error between the original and the compressed signal normalized to the maximum amplitude of the signal in the considered time window. Accordingly, $D(\cdot)$ takes value in $[0,1]$, and $D_\fl=1$.
\end{comment}
Notice that~\eqref{eq:dist} is a convex decreasing function of the compression ratio $k/m$.

\subsection{Outage probability} \label{subsec:outage}

We consider an erasure channel at the packet level, where the erasure probability depends on the rate $R(k)$ and the actual channel conditions. The communication channel is affected by both fading and Additive White Gaussian Noise (AWGN). We assume a quasi-static scenario, hence the fading coefficient $H$ remains constant over the packet duration. The device does not perform power control and the transmission power is fixed to $P_\ttx$; the Signal-to-Noise Ratio (SNR) $\gamma$ at the receiver is:~
\begin{equation}
 \gamma = \dfrac{|H|^2 P_\ttx}{A^2\, (d/d_0)^{\eta} N} \triangleq |H|^2 \, \bar{\gamma},
\end{equation}
where $N$ is the noise power, and the term $A^2\, (d/d_0)^{\eta}$ accounts for path loss and depends on the path-loss exponent $\eta$, the distance between transmitter and receiver $d$, and a path-loss coefficient $A=4\pi d_0 f_0/c$, where $f_0$ is the transmission frequency, $c$ the speed of light, and $d_0$ a reference distance for the antenna far field.

\begin{comment}
The communication channel is modeled as:~
\begin{equation}
 Y_i = H\,X_i + W_i \qquad i\in\mathbb{N},
\end{equation}

\noindent where $Y_i$ corresponds to the channel output, $X_i$ is the complex symbol transmitted over the $i$-th channel use, $W_i$ models the Additive White Gaussian Noise (AWGN)% with spectral density $N_0$
, and $H$ is the channel coefficient that represents fading and path loss, and is assumed to be constant over the packet duration.
%In the block fading model, the outage probability coincides with the erasure probability of a packet because all bits of the packet are equally affected.
When the channel is in a deep fade (i.e., $|H|$ is small), the packet is lost and a packet erasure (i.e., channel outage) occurs.
\end{comment}

When the channel is in a deep fade (i.e., $|H|$ is small), the packet is lost and a packet erasure (i.e., channel outage) occurs. 
In IoT scenarios where packets are likely to be short, the concepts of capacity and outage capacity of classic information theory are no longer applicable.
Recently, Polyansky, Poor and Verd\'u~\cite{polyansky} developed a finite-length information theory that revisits the classical concepts of capacity when packets have short size. In particular, they defined the \emph{maximum coding rate} $R^\star(S,\varepsilon)$ as the largest coding rate $L(k)/S$ for which there exists an encoder/decoder pair of packet length $S$ whose error probability is not larger than $\varepsilon$. Although the derivation of $R^\star(S,\varepsilon)$ is an NP-hard problem, tight bounds have been derived and, for quasi-static channels, the following approximation holds~\cite{yang}:~
\begin{equation}
 R^\star(S,\varepsilon) = C_\varepsilon + \mathcal{O}\left(\frac{\log_2 S}{S}\right),
\end{equation}

\noindent where $C_\varepsilon$ is the classical outage capacity for an error probability not larger than $\varepsilon$.
This is a very useful result, because it legitimates the use of the quantity $\log_2(1+\gamma)$ even in the finite-length regime~\cite{yang}.
Based on these considerations, we define the outage probability as:~
\begin{equation}
 \pr_\out(k) = \prr\left(\,\log_2\left(1+\gamma\right) < R(k)\right). \label{eq:out_gen}
\end{equation}

We assume the fading to follow a Rayleigh distribution, i.e., the fading matrix $H$ has i.i.d., zero-mean, unit-variance, complex Gaussian entries, and therefore the outage probability can be expressed as:~
\begin{equation} \label{eq:outage}
 \pr_\out(k) = 1 - e^{-(2^{R(k)}-1)/\bar{\gamma}},  
\end{equation}
The outage probability is non-decreasing in $R(k)$, and thus in $k$, being initially convex and then concave.
Notice that it increases with the distance $d$ from the receiver.

\subsection{Optimal rate-distortion point} \label{subsec:k_star}

Given the amount of energy available $u$ in the current slot, the device must decide how many bits of information $k$ to transmit within a packet of fixed size $S$. This corresponds to jointly selecting the source compression ratio and the coding rate.
The degree of compression employed affects the quality of the information by introducing a source distortion that is non-increasing in $k$, see Eq.~\eqref{eq:dist}.
On the other hand, the outage probability given in~\eqref{eq:outage} is non-decreasing in $k$, and thus there is a tradeoff between $D(k)$ and $\pr_\out(k)$.
The \emph{actual} distortion $\Delta(k)$ experienced by a packet at the receiver side is equal to $D(k)$ if the packet has been delivered, which happens with probability $1- \pr_\out(k)$, and is equal to $D_\fl$ otherwise.

Based on the definition of $\Delta(k)$, we set up RDP as follows:
\begin{subequations} \label{prob:FOP}
 \begin{flalign}
   \text{RDP:} && & k^\star = \min_{k\in \{0,\dots,m\}} \, \mathbb{E}[\Delta(k)] & \label{eq:RDP_objective}
 \end{flalign}
 \vspace{-\belowdisplayskip}
 \vspace{-\abovedisplayskip}
 \begin{alignat}{2}
   \shortintertext{subject to:}
    & E_{\rm used}(k) \le u &&  \label{eq:RDP_energy}
 \end{alignat}
\end{subequations}

\noindent where the expected \emph{overall} distortion at the receiver is:~
\begin{equation} \label{eq:overall_dist}
 \mathbb{E}[\Delta(k)] = D(k)\,(1- \pr_\out(k)) + D_\fl\,\pr_\out(k).
\end{equation}

The discrete nature of $k$ makes it harder to solve RDP. We thus introduce the continuous variable $w\in[0,m]$ and then map it into $k\in \{0,\dots,m\}$. Also, we initially analyze the energy and distortion aspects of the problem separately: we determine a value $k_E^\star$ that solves the energy constraint~\eqref{eq:RDP_energy} and a value $k_R^\star$ that solves RDP when Constraint~\eqref{eq:RDP_energy} is neglected, and then we combine them to obtain $k^\star$. 

We now focus solely on Constraint~\eqref{eq:RDP_energy}. When $u>0$, since $E_{\rm used}(\cdot)$ is non-decreasing in $w\in[0,m-1]$, there exists a point $w_E^\star(u)$ that minimizes the gap between the consumed and the allocated energy, which either solves Constraint~\eqref{eq:RDP_energy} with equality\footnote{If $E_P(w)$ is not strictly increasing in $w$, multiple rates may satisfy Constraint~\eqref{eq:RDP_energy} with equality. In this case, we select the maximum of these values.}, or is equal to $m-1$. Then, we choose $k_E^\star(u)$ as $\lfloor w_E^\star(u) \rfloor$. %largest $k \in \{0,\dots,m-1\}$ such that $k \le w_E^\star(u)$.
Instead, if no energy is allocated ($u=0$), then $k_E^\star(u) = 0$ and the packet is discarded. %\footnote{According to the energy consumption model, the sensing and circuitry costs in Eq.~\eqref{eq:e_circuitry} are constant quantities regardless of $k$. However, wthout loss of generality, we assume them not to be included in the energy allocation process.}. 
Finally, notice that $E_{\rm used}(m) < E_{\rm used}(k),\, k\in\{1,\dots,m\!-\!1\}$ because no energy is spent for the processing operations. We will see later how to include this case in the solution of RDP.

We now discuss how to determine $k_R^\star$, and present the following key result.

\begin{thm} \label{theorem:k_R}
    If $L_0\le S$, the expected overall distortion $\mathbb{E}[\Delta(w)]$ has exactly one local minimum at $w_R^\star$ in the continuous interval $[0,m]$.
    \begin{proof}
        See \appendixname~\ref{apx:proof_kR}.
    \end{proof}
\end{thm}
 
There exists no closed-form for $w_R^\star$ (see the computation in \appendixname~\ref{apx:proof_kR}); however, it is easy to numerically determine it by means of dichotomic search over a restricted subinterval of $[0,m]$ and then derive its discrete counterpart $k_R^\star$ accordingly. We refer the reader to \appendixname~\ref{apx:r_star} for details.

Based on these observations, the optimal $k^\star(u)$ is given by:~
\begin{equation} \label{eq:k_star}
 k^\star(u) = 
 \begin{cases}  
    \min\{k_R^\star, k_E^\star(u)\} & \text{if } k_R^\star<m \\
    m\cdot\chi_{\{E_{\rm used}(m)\le u\}} & \text{if }  k_R^\star=m \\
 \end{cases}
\end{equation}

Intuitively, the energy constraint~\eqref{eq:RDP_energy} does not allow to choose any $k\!>\!k_E^\star(u)$. On the other hand, it is not efficient to use any $k\!>\!k_R^\star$ even if possible, because it would lead to a worse expected distortion. When the joint source-channel coding optimization leads to $k_R^\star\!=\!m$, the packet is sent without being compressed, given that there is enough energy to perform the transmission.
Notice that, because of how $k^\star(u)$ is defined, the expected distortion at the receiver $\mathbb{E}[\Delta(k^\star(u))]$ is a non-increasing and convex function of $u$; see \appendixname~\ref{apx:proof_kR} for details about convexity.

\section{Energy Management} \label{sec:EMP}

Through Eq.~\eqref{eq:k_star}, RDP dictates the optimal coding rate $R^\star(u_n) \triangleq L(k^\star(u_n))/m$ when the energy $u_n$ available in slot $n$ is given. 
The allocation of $u_n$ is influenced by the dynamics of the energy inflow and by the battery state, which have been characterized in Section~\ref{subsec:eh}.
In this section, we formulate the distortion-energy problem as an MDP, which is solved by means of the well-known Value-Iteration Algorithm (VIA)~\cite{bertsekas}.

\subsection{Formulation of the Markov Decision Process} 
We model the problem as an MDP defined by the tuple $\left(\mathcal{S},\,\mathcal{U},\,P,\,c(\cdot)%,\,d(\cdot)
\right)$, with the components described as follows.
\begin{itemize}
 \item $\mathcal{S} \triangleq \mathcal{X} \times \mathcal{B}$ is the system state space, where $\mathcal{X}=\{0,1\}$ denotes the set of energy source states, and $\mathcal{B}=\{0,\dots,B\}$ represents the set of energy buffer states. %Both the energy source and buffer have been described in Section~\ref{subsec:eh}.
 
 \item $\mathcal{U}$ is the action set. In each slot, the device observes the current system state $s_n$ and chooses the amount of energy $u_n \in \mathcal{U}$ to use. This decision influences the joint source-channel coding scheme to adopt, as dictated by RDP through Eq.~\eqref{eq:k_star}. The set of admissible actions in state $s\in\mathcal{S}$ is $\mathcal{U}_s \subseteq \mathcal{U}$. According to Eq.~\eqref{eq:en_causality}, the usable energy is constrained by the present charge of the battery, thus $\mathcal{U}_{s_n} = \{0,\dots,b_n\}$.
 
 \item $P$ represents the transition probabilities that govern the system dynamics. %As described in Section~\ref{subsec:eh}, the source state is modeled as a 2-state MC, and the energy buffer evolves according to Eq.~\eqref{eq:buffer}. Thus, t
 In particular, the probability of going from state $s_n = (x_n,\, b_n)$ to $s_{n+1} = (x_{n+1},\, b_{n+1})$ when the action taken is $u_n$ is given by:~
 \begin{equation}
  \begin{split}
   \prr(s_{n+1}|s_n, u_n)=   &\; p_e (e_n|x_n) \cdot p_x (x_{n+1}|x_n) \\
   & \cdot \delta \left(b_{n+1} - \left(b_n + e_n - u_n \right)^\dagger \right)
  \end{split}
 \end{equation}
 
 \noindent where $p_x (x_{n+1}|x_n)$ is obtained from the transition probability matrix of the MC that models the source state,  $p_e (e_{n}|x_n)$ is the mass distribution function of the energy inflow in state $x_n$ (see Section~\ref{subsec:eh}), and $\delta(\cdot)$ is equal to one if the argument is zero, and zero otherwise. This last term is needed to ensure that the transitions between states are consistent with the dynamics of the battery state as determined by both the energy harvesting process and the transmission decisions.
 
 \item Finally, $c(\cdot)$ is the cost function. Because our goal is to minimize the distortion that affects the received packet, the cost in slot $n$ is a positive quantity defined as:~
 \begin{equation}
  c(s_n, u_n) = \mathbb{E}[\Delta(k^\star(u_n))],
 \end{equation}

 \noindent where $k^\star(u_n)$ is the result of the optimization of RDP. Notice that, apparently, the cost depends solely on the chosen action and not on the current state $s_n=(x_n,b_n)$. However, the action must be selected in set $\mathcal{U}_{s_n}$, which does depend on the present battery status $b_n$.
\end{itemize}

We aim at identifying a policy $\pi$, i.e., a sequence of decision rules that map the system state into the action to take.
As discussed in Section~\ref{sec:problem}, the goal is to minimize the long-term average distortion, i.e., the long-term average cost:~
\begin{equation}
 J^\pi = \lim\limits_{N\rightarrow +\infty} \frac{1}{N} \mathbb{E}_s\left[ \sum\limits_{n=0}^{N-1} c(s_n, u_n)\right], %\middle\rvert s_0 = s \right],
\end{equation}

\noindent which depends on the chosen admissible policy $\pi$ (that decides upon the action $u$). In general, it also depends on the initial state $s_0$, but the MDP we defined has a unichain structure and bounded costs, thus the asymptotic behavior is independent of the initial state and it is sufficient to consider only Markov policies~\cite{altman}, i.e., those for which the decision rule depends only on the current state $s_n$ and not on time $n$. 

The goal of EMP is to determine $J^\star = \min_{\pi\in \Phi} \, J^\pi$ and the corresponding optimal policy $\pi^\star = \argmin_{\pi\in \Phi} \, J^\pi$, where $\Phi$ denotes the set of all stationary Markov policies.

\begin{comment}
Note that the optimal control corresponding
to Eq. (29) is a pure policy whereby a single control u is associated with each state
; that is, there exists a mapping function μ(x) such that u(x) = μ(x) for each state
x ∈ X and u(x) is unique for each x.
\end{comment}

\subsection{Solution of EMP}
% TODO GUPTA 

\begin{comment}
$J^\star$ is proven to satisfy Bellman's optimality equation:~ 
\begin{equation} \label{eq:bellman}
 J(s) = \min_{u\in \mathcal{U}_s} \left\{c(s,u) + \sum\limits_{s^\prime\in\mathcal{S}} \prr(s^\prime|s, u)\,J(s^\prime)  \right\}
\end{equation}

\noindent \CP{cambiare notazione, chi e J(s)?)} where the immediate cost $c(s,u)$ obtained in the current state $s$ is summed with the expected optimal cost obtained from the next slot onwards, weighed according to the system dynamics.
It can be solved, e.g., through the Relative Value-Iteration Algorithm (RVIA), a version of the VIA used for infinite-horizon average cost MDPs~\cite{bertsekas}. RVIA starts from an initial estimate of $J(s)\:\forall s\in\mathcal{S}$, and then uses Eq.~\eqref{eq:bellman} as the update rule. %TODO vedi Aprem pag 899
In particular, the $n$-th iteration of RVIA determines:~
\end{comment}

$J^\star$ can be proved to satisfy Bellman's optimality equation~\cite{bertsekas}, and thus our MDP can be solved, e.g., through the Relative Value-Iteration Algorithm (RVIA), a version of the VIA used for infinite-horizon average cost MDPs~\cite{bertsekas}. RVIA defines two functions $J$ and $Q$ that are iteratively updated starting from an initial estimate $J_0(\cdot)$, e.g., $J _0(s)=0\;\forall s\!\in\!\mathcal{S}$.
In particular, the $i$-th iteration determines:~
\begin{align}
  & Q_{i}(s,u) = c(s,u) + \sum\limits_{s^\prime\in\mathcal{S}} \prr(s^\prime|s, u)\,J_{i-1}(s^\prime) \label{eq:Q}\\
  & J_{i}(s) = \min_{u\in \mathcal{U}_s} Q_{i}(s,u), \label{eq:J_Q}
 \end{align}
 \noindent 
 In~\eqref{eq:Q}, the immediate cost $c(s,u)$ obtained in the current state $s$ is summed with the expected optimal cost obtained from the next slot onwards, weighed according to the system dynamics.
The convergence criterion is given by the span seminorm operator $sp(j) \triangleq  \max(j)-\min(j)$ computed for $j = J_{i+1}(s) - J_i(s)$; the span seminorm guarantees that~\eqref{eq:J_Q} %Bellman's optimality equation
 is a contraction mapping, and thus the RVIA algorithm is assured to converge (for details about convergence, we refer the reader to~\cite{bertsekas}). 
We stop the iterative algorithm when $sp(\cdot)\le \varepsilon$, for a chosen tolerance threshold $\varepsilon$.   
The optimal policy $\pi^\star$ is then determined by computing the optimal action to follow in each state $s\in\mathcal{S}$, i.e.,  $u^\star(s) = \argmin_{u\in \mathcal{U}_s} Q_{n}(s,u)$, where $n$ is the last iteration of RVIA, and has the following key property. % of the structure of the optimal policy.

\begin{thm}\label{theorem:EMP}
 The optimal policy is non-decreasing in the energy buffer state $b$.
  \begin{proof}
    See \appendixname~\ref{apx:proof_EMP}.
  \end{proof}
\end{thm}

By taking into account the discrete nature of the system state, we have that the optimal policy has a threshold structure, which is an appealing property for storage and implementation on resource constrained nodes.

When the optimal policy is identified, the MDP turns into a MC, and the average long-term cost induced by $\pi^{\star}$ is:~
\begin{equation}
 C_\pi^\star =\sum\limits_{s\in\mathcal{S}} \rho_s\, c(s,u^\star(s)) \label{eq:final_cost}
\end{equation}

\noindent where $\rho_s$ is the steady state probability of state $s$ induced by the optimal policy, and $c(s,u^\star(s))$ is the immediate cost obtained by following $\pi^\star$ in state $s$.

\smallskip
\textit{\textbf{Note on computational complexity:}}
We highlight that, although the determination of the optimal policy requires to solve the MDP, the RVIA algorithm does not need to be computed at runtime. As discussed, e.g, in~\cite{bui, Aprem}, the policy can be obtained offline and not necessarily by the sensor nodes, which will just need to store tables containing the association between system state and optimal action. Therefore, the policy execution simply consists in a look-up operation.
%TODO in Bui mettono valori con gli O grandi sulla complexity

% \section{Packet losses} \label{sec:retx}

\section{Numerical evaluation} \label{sec:results}
%\CP{forse servirebbe grafico numero medio di slot in cui tx}

\begin{figure}[t]
  \centering
  \includegraphics[width=\columnwidth]{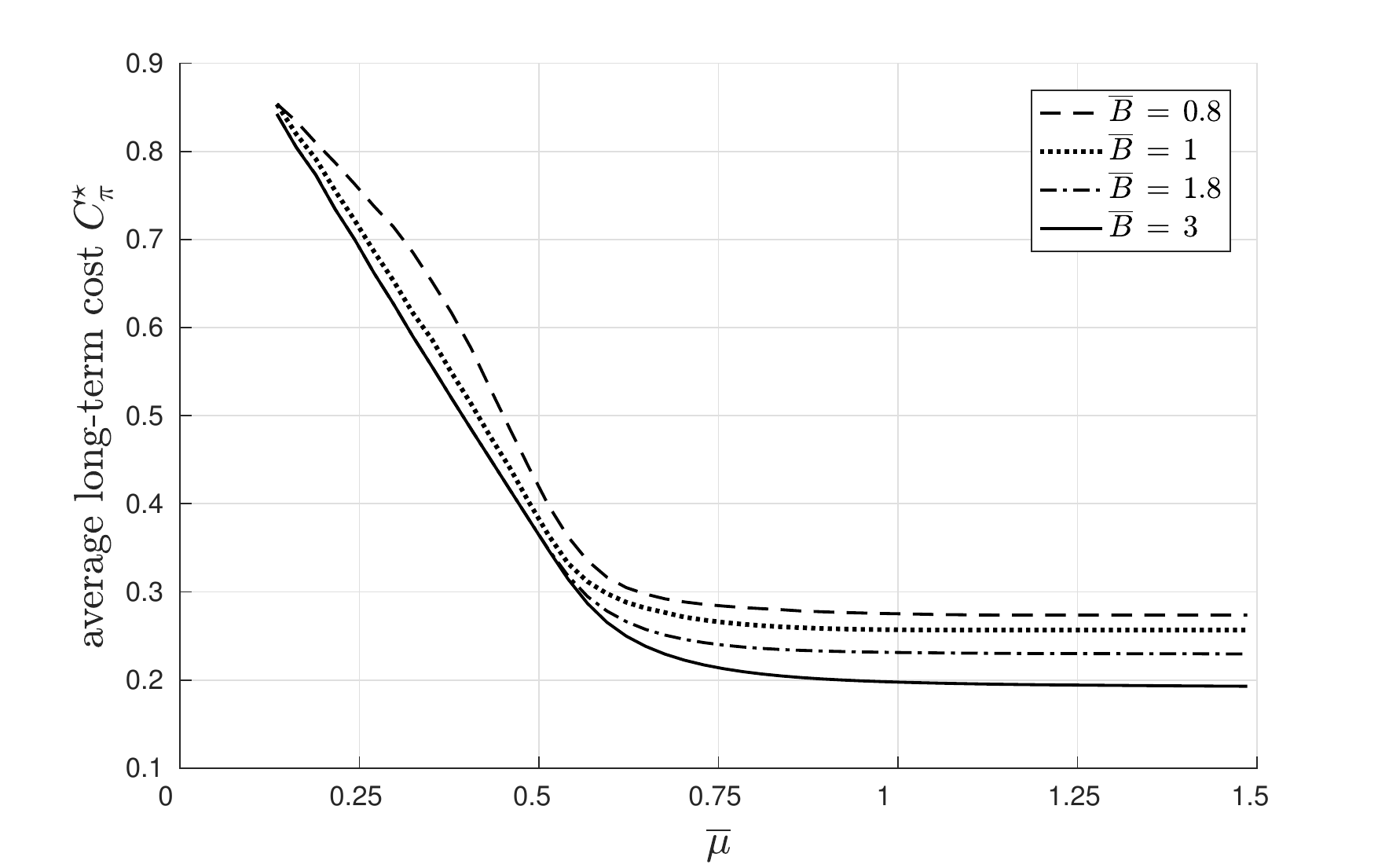}
  \caption{$C_\pi^\star$ vs $\overline{\mu}$ for different values of $\overline{B}$ ($d=100$ m).}
  \label{fig:B_mu}
\end{figure}

% settare parametri.
% Vedere effetto dei vari params:
% \begin{itemize}
% \item canale (-> Poutage) // dovrebbe essere che aumentando la distanza, peggioro il canale e quindi distorsione piu bassa
% \item num livelli compressione: NO. o cambio anche consumo en, o non ha senso perche semplicemente aumento E spesa e quindi il costo mi aumenta. PARAM m VA TENUTO FISSO
% \item batteria // Claramentri, pi\'u alta la batteria, pi\'u basso \'e il costo perch\'e posso raggiungere kR pi\'u facilmente. Tuttavia, dopo un certo punto satura.
% \item param energetici ($E, \mu_E, \sigma_E^2$): UPDATE: cambio mu_E (ha piu senso che E in effetti) e mean_time1
% \end{itemize}

To validate the analytical results, we evaluated the performance of the optimal policy  by investigating the effect of some parameters and through comparison with two heuristics. 

Concerning the processing aspects, we set $L_0=S$, $a=0.69$, $b=3.27$, and $m=20$.
The channel gain is computed using the standard path loss model with a path loss exponent equal to $3.5$ and a central frequency of $868.3$~MHz. The bandwidth is $W = 125$~kHz, and the overall noise power is $-167$~dBm/Hz. 
For the energy aspect, the probability that the source goes from the bad to the good state is 3 times greater than that of the opposite transition. We set the variance of the normal distributed energy arrivals in the ``good'' state $\sigma^2=3$, whereas the mean $\mu$ will be specified for each result. %, and the maximum number of quanta that can be harvested in a slot is $E=\mu/0.65$.

\begin{figure}[t]
  \centering
  \includegraphics[width=\columnwidth]{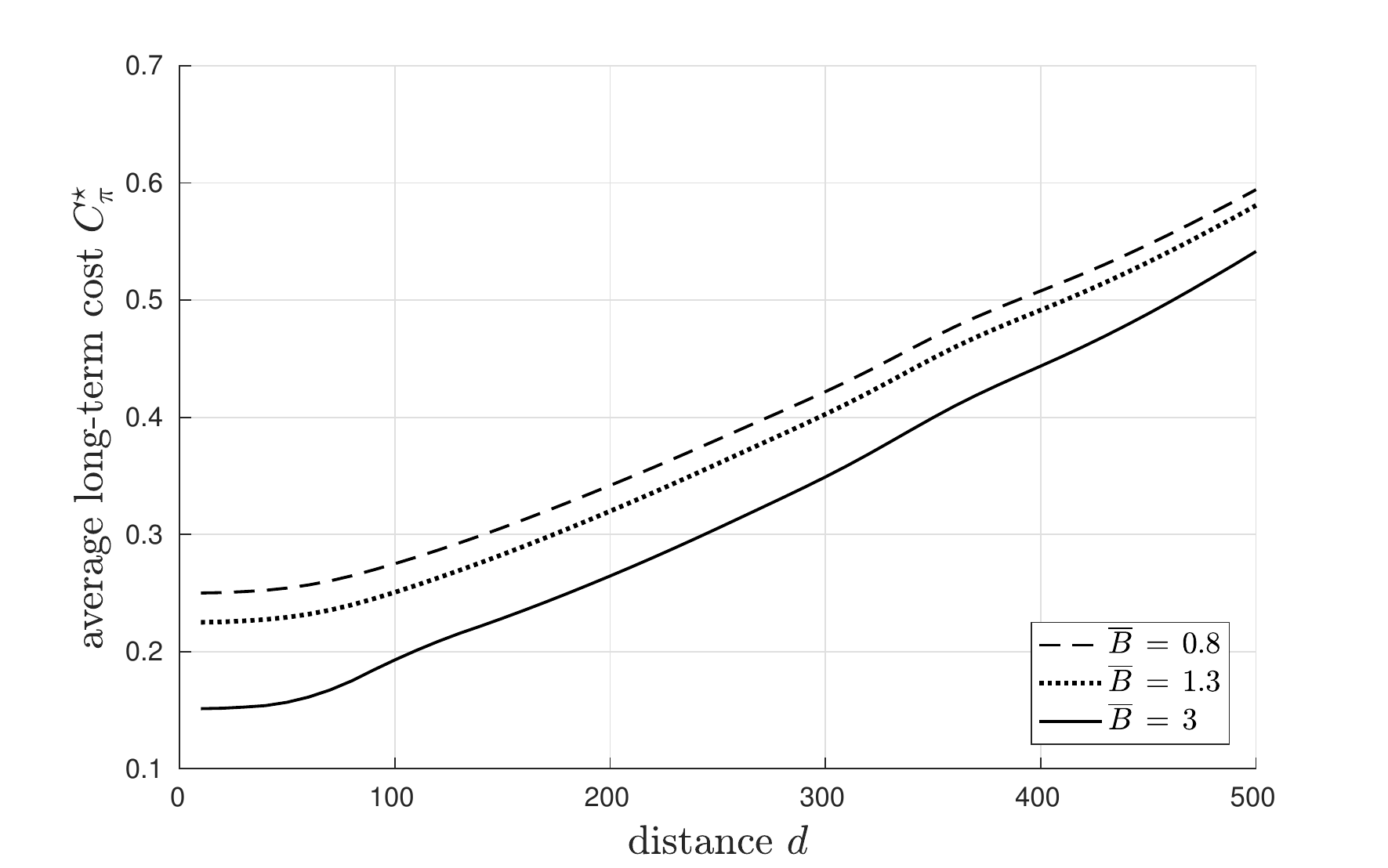}
  \caption{$C_\pi^\star$ vs $d$ for different values of $\overline{B}$ ($\overline{\mu}=1$).}
  \label{fig:B_dist}
\end{figure}

We introduce the following useful quantities: $e_{\max}$ is the maximum energy consumption demanded by the processing and transmission of a packet, 
$\overline{\mu} = \mu/e_{\max}$, and $\overline{B}=B/e_{\max}$. We remark that, according to~\eqref{eq:energy_used}, a packet cannot be sent if the available energy is below a certain threshold $e_{\min}$. %; in the simulations, we have $e_{\min} > e_{\max}/2$, which implies that 

Fig.~\ref{fig:B_mu} shows the average long-term cost $C_\pi^\star$ as a function of $\overline{\mu}$ for different values of $\overline{B}$, when the distance is fixed to 100 m. Clearly, the more the energy that can be harvested, the better the performance. It is interesting to note that, as long as it is sufficiently large, the battery size has only a minor impact on the average cost: when it is large enough to allow that $k_R^\star$ is reached in most of the time slots, it is not worth to further increase $B$, because the exceeding energy is not used.
Notice that, after a certain value of $\overline{\mu}$ (which depends on the other parameters), the average cost tends to stabilize around a specific configuration-dependent value. The reason behind it is twofold: (i) the distortion associated to $k_R^\star$ is never zero, unless the outage probability is negligible even for $k=m$ \footnote{We discarded this scenario, as it is not interesting. For $k<m$, it is always $\Delta(\cdot)\ge D(\cdot)>0$.}, and (ii) a minimum amount of energy is needed to transmit a packet and therefore, for all system states where the battery is $b<e_{\min}$, the cost is 0, and this has an impact on $C^\star$, see Eq.~\eqref{eq:final_cost}.

%This minimum achievable distortion is reached for $k=k_R^\star$ (see Section~\ref{sec:RDP}) and is greater than zero because $D$ and $P_{\out}$ are both positive quantities.

Fig.~\ref{fig:B_dist} evaluates the performance of the optimal policy against the distance from the receiver for different values of the battery size when $\overline{\mu} = 1$. As $d$ increases, the channel becomes worse and the higher outage probability has a negative impact on the achievable distortion at the receiver. Again, the role of $B$ is less relevant.

\subsection{Comparison with heuristics} \label{subsec:eu}
% \textbf{\textit{Comparison with heuristics:}}

We also compare the performance of the optimal policy (OP) against that of the two following heuristic policies.
\begin{itemize}
 \item \emph{Greedy policy (GP)}: the future sustainability of the node does not influence the energy to use in each slot, which is chosen as $u_n= \min(b_n, u^\star)$, with $u^\star$ being the energy needed to achieve $k_R^\star$. Basically, GP only solves RDP and does not optimize the energy utilization.
 
 \item \emph{Dumb processing policy (DP)}: it does not determine the optimal point in the rate-distortion tradeoff and only considers the distortion introduced at the source, without accounting for the outage probability. EMP is unchanged. %and only aims at guaranteeing the node's self-sufficiency. In practice, it is the opposite of GP and does not solve RDP.
\end{itemize}

\begin{figure}[t]
  \centering
  \includegraphics[width=\columnwidth]{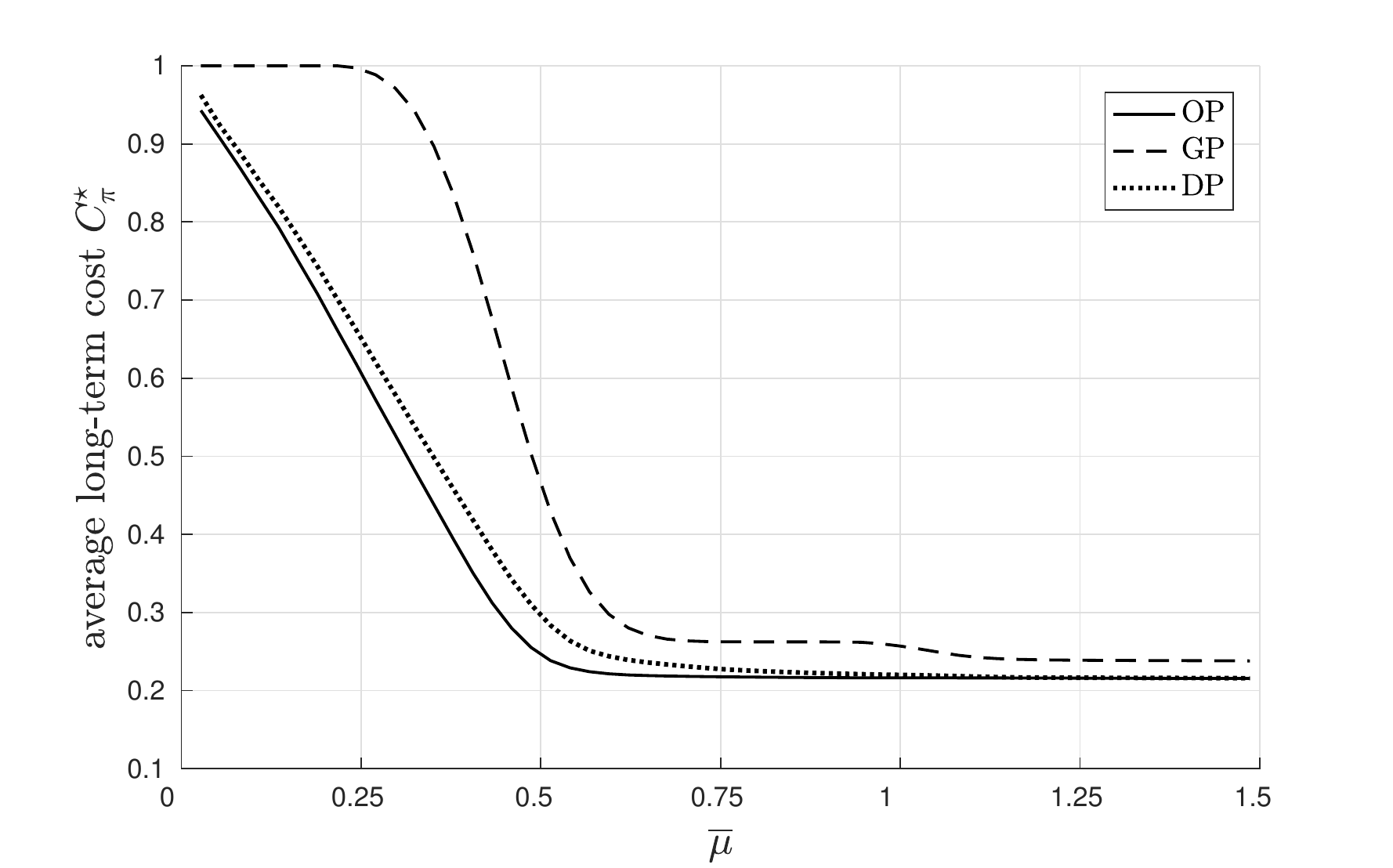}
  \caption{Performance comparison between policies.}%: $C_\pi^\star$ v. $\overline{\mu}$ ($\overline{B}=1.5$, $d=80$ m).}
  \label{fig:heur}
\end{figure}

Fig.~\ref{fig:heur} compares the performance of the optimal and heuristic policies against the average energy income during the good state of the energy source. The normalized battery size is $\overline{B} = 1.5$ and the distance from the receiver is $d=80$ m. As expected, the optimal policy outperforms both heuristic policies. However, DP achieves almost optimal performance with this configuration: this is explained by the fact that the source distortion becomes flat as $k$ increases (see~\eqref{eq:dist}), thus the gain brought by OP for considering the negative impact of the outage probability is less significant than that of the intelligent energy management process\footnote{Accordingly, if the outage were higher, the gap between DP and OP would increase.}. In fact GP, that makes an aggressive use of the available energy, leads to the worst performance. % because it generally does not allow to accumulate enough energy to obtain a low distortion.
When the average energy income $\mu$ is high, all policies achieve good performance, because there is almost always enough energy to guarantee low distortion. % even for GP.

\begin{figure}[t]
  \centering
  \includegraphics[width=\columnwidth]{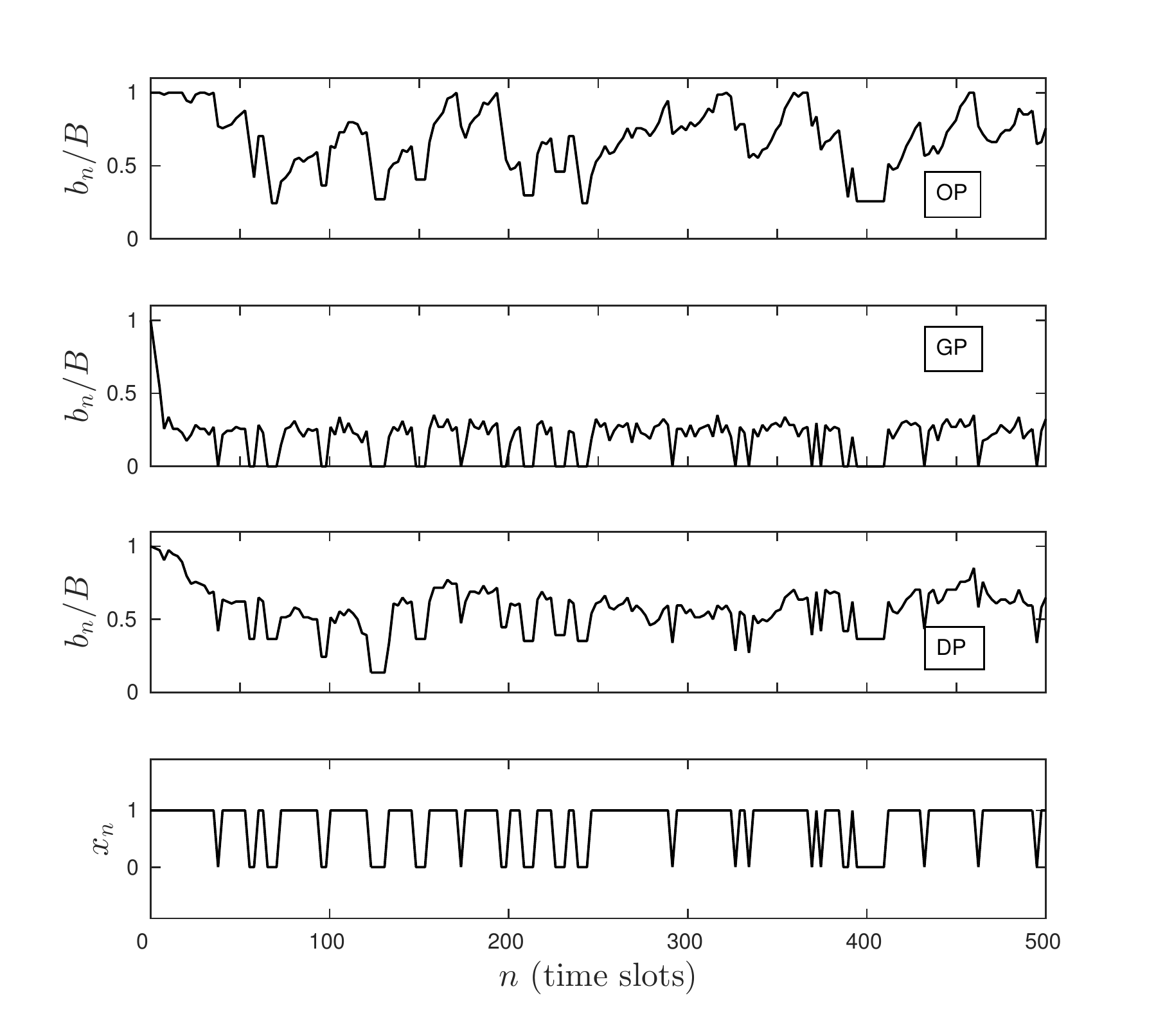}
  \caption{Example of battery temporal evolution ($\overline{B} = 2$).}
  \label{fig:en_dyn}
\end{figure}

In Fig.~\ref{fig:en_dyn}, we plot a realization of the system temporal evolution during the first $500$ time slots, when $d = 80$ m and starting with full battery ($\overline{B} = 2$).
The last graph shows the dynamics of the energy source, while the first three plots show the evolution of the energy buffer for OP, GP and DP, respectively. %When using GP, the battery is always half empty, because the policy does not permit to save energy for future slots and the battery charge is only determined by the external energy income.
GP drains out the battery too fast, and the battery is often empty when $x_n=0$, whereas this never happens with OP and DP, which leverage on the knowledge of the EH statistics to use the energy wisely. Interestingly, the excursion experienced by $b_n$ is lower with DP than with OP: it neglects the negative impact of the outage probability for large values of $k$ and tends to use energy to send the packet even when OP decides to discard it, and in this way it does not allow the buffer to accumulate energy that may be very useful in the next slots.

%tends to use more energy than OP when available, because it neglects the negative impact of the outage probability for large values of $k$, and in this way it does not allow the buffer to accumulate energy that may be very useful in the next slots.

\section{Conclusions} \label{sec:conclu}

% modularity
This work investigates the tradeoff between energy efficiency and signal distortion at the receiver. 
We decomposed the problem into two nested optimization steps, endowing our framework with more flexibility. The outer problem is structured as an MDP, that allows to derive an optimal energy management policy; the inner problem jointly optimizes the source-channel coding scheme to ensure that the quality of the received information is maximized.
The numerical evaluation corroborates the analytical results and shows that our policy outperforms simpler heuristics. 

Future work includes the use of machine learning techniques in order to more accurately model the EH process and the extension to a sensor network where packet re-transmissions are allowed, thereby introducing a new tradeoff between signal distortion and transmission latency. We also would like to address more general cases, e.g., considering compression algorithms that behave differently from LTC in terms of energy consumption, and including practical sensor limitations, such as imprecisions in the energy buffer readings.

\section*{Acknowledgments}
The work of C. Pielli and M. Zorzi was partly supported by Intel's Corporate Research Council, under the project ``EC-CENTRIC: an energy- and context-centric optimization framework for IoT nodes''. The work of \v C. Stefanovi\' c was supported by the Danish Council for Independent Research under grant no. DFF-4005-00281.

\appendices

\section{Proof of Theorem~\ref{theorem:k_R}}\label{apx:proof_kR}

By substituting $D(\cdot)$ and $\pr_\out(\cdot)$ in Eq.~\eqref{eq:RDP_objective} as given in Eqs.~\eqref{eq:dist} and~\eqref{eq:outage}, respectively, the expected distortion for $w\in[1,m]$ is expressed as:~
\begin{equation}
\begin{split}
 \mathbb{E}[\Delta(w)] & =\: b\left(\left(\frac{w}{m}\right)^{-a} - 1\right) \, e^{-\frac{2^{(w L_0)/(m S)}-1}{\gamma}} \\
 & + \: D_\fl\,\left(1-e^{-\frac{2^{(w L_0)/(m S)}-1}{\gamma}}\right) \le D_\fl, \label{eq:delta}
\end{split}
\end{equation}

\noindent whereas it is equal to $D_\fl$ for $w=0$. By introducing the positive coefficients $c_1, c_2, c_3, c_4$ (given in Table~\ref{table:coeff}), Eq.~\eqref{eq:delta} can be rewritten as:~
\begin{equation} 
 \mathbb{E}[\Delta(w)] = c_1\,w^{-a}\,e^{-c_2\,2^{w\,c_3}} - c_4\,e^{-c_2\,2^{w\,c_3}} + D_\fl,
\end{equation}

\noindent and its derivative with respect to $w$ is:~
\begin{equation}
\begin{split}
 \frac{\partial}{\partial w} \mathbb{E}[\Delta(w)] & = -e^{-c_2 2^{w c_3}} \left(d_1 w^{-a-1}\!+\!d_2 w^{-a} 2^{w c_3}\!-\!d_3 2^{w c_3} \right) \\
 & \triangleq - e^{-c_2\,2^{w\,c_3}} \, f(w), \label{eq:f_der}
\end{split}
\end{equation}

\noindent with positive coefficients $d_1, d_2, d_3$ detailed in Table~\ref{table:coeff}. 
The critical points of $\mathbb{E}[\Delta(w)]$ correspond to the zeros of $f(w)$. However, it is not possible to write a closed-form solution for $f(w)=0$, but a numerical approach needs to be pursued.
\begin{table}[t]
    \centering	
    \begin{tabular}{ c | c | c | c }	    
	    \toprule	
	    $c_1$	& $c_2$	 &  $c_3$ & $c_4$		\\
	    \hline
	    $b \,m^{a}e^{1/\gamma}$ & $1/\gamma$ & $L_0 / (m\,S)$ & $e^{1/\gamma} (b+D_\fl)$ 	\\
	    \bottomrule
    \end{tabular}
    
    \medskip
    
    \begin{tabular}{ c | c | c | c }	    
	    \toprule	
	     $d_1$ & $d_2$  &  $d_3$ & $d_4$ \\
	     \hline	
	     $a\,c_1$ & $c_1 c_2 c_3 \ln(2)$  &  $c_2 c_3 c_4 \ln(2)$ & $e^{1/\gamma} c_2 c_3 \ln(2)$	\\
	    \bottomrule
    \end{tabular}
    \caption{Coefficients  $c_1, c_2, c_3, c_4,d_1, d_2, d_3$.} 
    \label{table:coeff}
\end{table}

Actually, the shape of $f(w)$ highly depends on coefficients $d_1,d_2, d_3$, and thus on the system parameters.
Anyway, we show that, if $L_0\le S$, $f(w)$ has at most one root in the continuous interval $[1,m]$, and thus function $\mathbb{E}[\Delta(w)]$ admits a unique point of minimum $w_R^\star$ (which may coincide with one of the extreme values of $w$).

We temporarily focus only on the last two addends of $f(w)$ and define $g(w) \triangleq (d_2\,w^{-a} - d_3)\,2^{w\,c_3}$. By expanding coefficients $d_2$ and $d_3$, we obtain:~
\begin{equation}
 g(w) =  d_4\,\left(b \,\left(\frac{m}{w}\right)^{a} - (b + D_\fl)\right)\,2^{w\,c_3},
\end{equation}
\noindent with $d_4$ defined in Table~\ref{table:coeff}.
The derivative is:~
\begin{equation}
\begin{split}
 \frac{\partial}{\partial w}& g(w) = d_4\,2^{w\,c_3} \Big(-a\,b\,m^a\,w^{-a-1} +\\
 &+ b \left(\frac{m}{w}\right)^a \ln(2) c_3 - b \ln(2) c_3 - D_\fl \ln(2) c_3\Big)
 \end{split}
\end{equation}

The last addend of the second factor is certainly negative, and the sum of the first three addends is equal to:~
\begin{equation}
\begin{split}
 &b\,\left(\frac{m}{w}\right)^a \left(-\frac{a}{w} + \ln(2)c_3\left(1- \left(\frac{w}{m}\right)^a\right) \right) \le \\
 &\le b\,\left(\frac{m}{w}\right)^a \left(-\frac{a}{w} + \frac{1}{m}\left(1- \left(\frac{w}{m}\right)^a\right) \right)  \triangleq b\,\frac{m^a}{w^a} h(w)
 \end{split}
\end{equation}
 \noindent where the first inequality comes from the assumption $L_0\le S$, that implies $c_3\le 1/m$. Notice that both $h(0)$ and $h(m)$ are negative, and $h'(w)$ is positive, i.e., function $h(\cdot)$ is a monotonically increasing negative function of  $w\in[0,1]$. It follows that, for $w\le m$, $g(w)$ is always decreasing in $w$. Moreover, it is certainly positive for $w\rightarrow0^+$ and negative for $w=m$. This means that $g(w)$ has exactly one point of minimum $w_g = (b/(b+D_\fl))^{1/a}$ in the continuous range $[0,m]$. However, if $w_g <1$, then $g(w)$ is always negative in the interval we are interested in (i.e., $w \in[1,m]$).

%\CP{allora. In realtà il termine dg2+dg3 e sempre positivo, quindi il ragionamento e sbagliato. considerando pero dg1+dg2+dg3, se c3 e piccolo (in particolare, L0<S/ln(2), ma io dico solo L0 < S e giustifico che ha senso) allora, ho $(m/k)^\alpha (-a/k + ln(2)c_3(1-(k/m)^\alpha)$ che e sempre<0,perche sostituisco ln(2)c3 con 1, e allora e vero che g e sempre negativa}

Based on these observations and on the fact that $d_1\,w^{-a-1}$ is positive, convex and decreasing in $w$, we obtain that $f(w)$ is decreasing in $w$. Depending on the values of $d_1, d_2, d_3$, it is then possible to pinpoint three distinct cases.

 \begin{enumerate}
  \item \label{enum:pos} $f(w)$ is always positive, and thus $\mathbb{E}[\Delta(w)]$ is always increasing in $w$, which entails that choosing $w_R^\star = 1$ is optimal. This happens when coefficient $d_1$ is very high and thus the first addend of $f(w)$ always dominates $g(w)$, i.e., when the outage probability is overwhelming. We deem this case not to be of practical interest.
  
  \item \label{enum:neg} In contrast to the previous case, it may happen that $f(w)$ is always negative, i.e., $\mathbb{E}[\Delta(w)]$ always decreases with $w$. This situation is met when the distortion function prevails over the outage probability, which remains low even for high rates because the channel is very good on average. The solution in this case is $w_R^\star=m$, i.e., the best strategy consists in not compressing at all. 
  
  \item \label{enum:int} Otherwise, $f(w)$ is decreasing in $w$ and has a unique zero, which corresponds to a unique value $w_R^\star$. Using a source compression ratio lower than $w_R^\star/m$ would generate a higher  $\mathbb{E}[\Delta(\cdot)]$ because of the larger distortion introduced with the source coding, whereas a compression ratio larger than $w_R^\star/m$ would weaken the goodness of the channel coding and lead to a higher outage probability.
 \end{enumerate}
 
In practice, under the only assumption of $L_0 \le S$, there always exists a unique point $w_R^\star$ that minimizes the expected distortion at the receiver, i.e., $\mathbb{E}[\Delta(w)]$ is decreasing until $w_R^\star$ and increasing afterwards. 
%Notice that it never is $k_{\min}=0$, since discarding the packet always leads to the highest distortion level.
\qed

\smallskip
\textbf{ \textit{A note on convexity.}}
$\mathbb{E}[\Delta(w)]$ is \emph{convex} for $w\le w_R^\star$.
In fact, the computation of the second derivative of~\eqref{eq:delta} leads to:
\begin{equation}
\begin{split}
  \frac{\partial^2}{\partial w^2} \mathbb{E}[\Delta(w)] & =  c_2 c_3 \ln(2) 2^{w c_3} e^{-c_2 2^{w\,c_3}}\,f(w) + \\
  & - c_2 e^{-c_2 2^{w\,c_3}}\,\frac{\partial}{\partial w}f(w) 
  \end{split}
\end{equation}
which is always positive for $w\le w_R^\star$, since $f(w)$ is positive and its derivative $-(a+1) d_1 w^{-a-2} +  \frac{\partial}{\partial w} g(w)$ is always negative.

\section{On the determination of $k_R^\star$} \label{apx:r_star}
The computation in \appendixname~\ref{apx:proof_kR} shows that $w_R^\star$ cannot be expressed with a closed-form. So, in the worst case $\mathbb{E}[\Delta(k)]$ should be calculated for all the $m$ possible values of $k$. 
However, it is still possible to determine $k_R^\star$ with the following procedure, that reduces the computational time. 
First of all, we compute the value of Eq.~\eqref{eq:delta} for the two extreme values $k\in\{1,m\}$. If both $f(1)$ and $f(m)$ are positive, then we fall in case~\ref{enum:pos}) of \appendixname~\ref{apx:proof_kR} and $k_R^\star = 1$; similarly, if they are both negative, case~\ref{enum:neg}) holds and $k_R^\star = m$. If $f(1)>0$ and $f(m)<0$, case~\ref{enum:int}) holds and $k_R^\star$ can be found by means of binary search over the discrete subinterval $[k_g, m-1]$, where $k_g$ is one of the two closest integers to $w_g$, according to which one yields the lowest expected distortion.

%. Then $k_R^\star$ is one of the two closest integers to $w_R^\star$, according to which one yields the lowest expected distortion.

\section{Proof of Theorem~\ref{theorem:EMP}}\label{apx:proof_EMP}
\begin{comment}
 TOPKIS:
 $f:X\times R \rightarrow R$ supermodular e  $x^{\star}(t) = argmax_{x\in S(t)} f(x,t)$. If $t'\ge t, S(t) \ge S(t')$, then $x^\star (t) \ge x^\star(t')$.
 
 Se g = -f \`e submodular, allora f \'e supermodular e argmin g = argmax -f. \\ 
 Quindi f = -Q, supermodular TIC\\
 t = u, S(t) = Us(u)\\
 Se u' > u, allora Us(u')>Us(u) TIC\\
\end{comment}

Thanks to Topkis' monotonicity theorem~\cite{topkis}, to prove that $u^{\star}(s) = \argmin_{u\in\mathcal{U}_s} Q((x,b),u)$ is non-decreasing in $b$, it is sufficient to prove that function $Q((x,b),u)$ is \emph{submodular} in $(b,u)$, i.e., that the difference $Q((x,b),u') - Q((x,b),u)$ with $u'\ge u$ does not increase as $b$ increases:~
\begin{equation}
 Q((x,b'),u') - Q((x,b'),u) \le Q((x,b),u') - Q((x,b),u)
\end{equation}
\noindent for $b'\ge b$.
The submodularity property is preserved under nonnegative linear combination, thus we separately demonstrate the submodularity of $c(s,u)$ and $J(s)$ in the pair $(b,u)$. In fact, from Eq.~\eqref{eq:J_Q}, we have that $Q((x,b),u) = c((x,b),u) + \sum_{x^\prime\in\mathcal{X}}\sum_{e=0}^E p_x(x'|x) p_e(e|x)  \,J((x^\prime, b+e-u))$.

\smallskip
For what concerns the single-stage cost $c((x,b),u)$, we have seen in Section~\ref{subsec:k_star} that it is convex and non-increasing in $u$, and depends on $b$ only through the set of admissible actions $\mathcal{U}_{(x,b)}$. Since $\mathcal{U}_{(x,b)} \subseteq \mathcal{U}_{(x,b')}$ if $b\le b'$, it is %always $c((x,b),u) \le c((x,b'),u)$
$c((x,b),u) < c((x,b'),u)$ for $b<u\le b'$, and $c((x,b),u) = c((x,b'),u)$ in all other cases. It follows that the quantity $c((x,b'),u) - c((x,b),u)$ does not increase as $u$ increases, i.e., the single-stage cost is submodular in $(b,u)$. 

We now proceed with the characterization of the term $J((x, b))$.
We assume it to be convex in the energy buffer state $b$; the validity of this property is proven in Lemma~\ref{lemma:J_convex} below.
Convexity means that:~
\begin{equation} \label{eq:J_convex}
\begin{split}
 J((x, b_1)) + J((x, b_2)) & \ge  J((x, \lambda b_1 + (1-\lambda)b_2)) +\\
 & + J((x, (1-\lambda)b_1 + \lambda b_2))
\end{split}
\end{equation}

\noindent for any $\lambda \in [0,1],b_1,b_2,x$. By choosing $b_1=b+e-u'$, $b_2=b'+e-u$, and $\lambda=(u'-u)/(u'-u+b'-b)$ with $b'\ge b$ and $u'\ge u$, and by rearranging the terms, we obtain:~
\begin{equation} \label{eq:J_submodular}
\begin{split} 
 J((x, b'+e-&u')) - J((x, b'+e-u)) \le \\
  & J((x, b+e- u')) - J((x, b+e-u)),
\end{split}
\end{equation}

\noindent which proves the submodularity of $J(x,b)$ in $(b,u)$.

Summing up, by exploiting the submodularity of function $Q((x,b),u)$ and the convexity of $J((x,b))$ with respect to $b$, we have proved that the optimal policy $\pi^{\star}(s)$ is non-decreasing in the buffer state $b$.

\begin{lemma} \label{lemma:J_convex}
%The expected long-term average cost 
$J(s)$ is convex in the energy buffer state $b$.
\end{lemma} %and non-increasing 

\noindent \emph{Proof.} We proceed by induction on the $i$-th iteration of RVIA (see Eq.~\eqref{eq:J_Q}).
 
  The basis is straightforward: since the initial value $J_0((x, b))$ does not affect the convergence of the algorithm, we simply choose it convex in $b$ (e.g., $J_0((x, b))=0 \:\forall s\in\mathcal{S}$ is a valid choice).
  
  For the inductive step, we assume $J_{i-1}((x, b))$ to be convex in $b$.  In the proof of Theorem~\ref{theorem:EMP} we have shown that $Q(s,u)$ is submodular in $b$ if $J((x, b))$ is convex in $b$, i.e.:~  
  \begin{equation} \label{eq:Q_submodular}
  \begin{split}
   Q_i((x,b),u') - & Q_i((x,b-c),u') \le \\
   & Q_i((x,b),u) - Q_i((x,b-c),u), 
  \end{split}
  \end{equation}
  
  \noindent for $b'\!\ge\!b$, $u'\!\ge\!u$.
  Function $Q_i((x,b),u)$ is convex in $b$, because it is the nonnegative weighted sum of $c((x,b),u)$ and $J_{i-1}((x, b))$, that are both convex in $b$. Thus:~
    \begin{equation}
  \begin{split}
   Q_i((x,b),u) - & Q_i((x,b-c),u) \le \\
   & Q_i((x,b+c),u) - Q_i((x,b),u), 
  \end{split}
  \end{equation}
  
  \noindent which, combined with~\eqref{eq:Q_submodular}, gives:~  
  \begin{equation}
  \begin{split}
    Q_i((x,b),u') - & Q_i((x,b-c),u') \le \\
   & Q_i((x,b+c),u) - Q_i((x,b),u),
  \end{split}
  \end{equation}  
\vspace{-0.41cm}
  \begin{equation}
  \begin{split}
    Q_i((x,b,u) - & Q_i((x,b-c),u') \le \\
   & Q_i((x,b+c),u) - Q_i((x,b),u').
  \end{split}
  \end{equation}
  
  \noindent By choosing $u=u'=\argmin_{\underline{u}} Q_i((x,b),\underline{u})$, we obtain:~  
   \begin{equation}
    J_i((x,b+c)) - J_i((x,b)) \ge    J_i((x,b)) - J_i((x,b-c)),
  \end{equation}
  
  \noindent i.e., $J_i(s)$ is convex in $b$ and the inductive step is proved.

\bibliographystyle{IEEEtran}


\end{document}